\documentclass [11pt, article] {article}	
\usepackage[margin=1in]{geometry}	
\usepackage[utf8]{inputenc}
\usepackage{amsmath}
\usepackage{amssymb}
\usepackage{array}

\usepackage{array,booktabs}
\usepackage{multirow}
\usepackage{graphicx}
\usepackage{booktabs}
\usepackage{float}
\usepackage{color,soul}
\usepackage{setspace}
\usepackage{subcaption}
\usepackage{siunitx}

\usepackage[style=numeric,backend=biber,sorting=none]{biblatex}
\usepackage[export]{adjustbox}
\usepackage{textcomp}

\DeclareGraphicsExtensions{.png,.pdf,.eps}
\title{Analysis of melting and flow in the hot-end of a material extrusion 3D printer using X-ray computed tomography}

\author{Julian Kattinger$^{*}$, Mike Kornely, Julian Ehrler, Christian Bonten and Marc Kreutzbruck\\
\small Institut für Kunststofftechnik \\
\small University of Stuttgart, Stuttgart, Germany\\
\small * Corresponding author}

\date{}

\begin{document}

\maketitle

\begin{abstract}
This paper presents in-situ X-ray computed tomography (CT) experiments used to study the flow behavior within a conventional hot-end of a fused filament fabrication printer. Three types of experiments were performed to better understand the melt and flow behavior. In one experiment, 360° CT scans were conducted, focusing on the air gap between the filament and the nozzle wall. In a second experiment, the flow profile inside the nozzle was studied using radiography. To provide a good contrast to the surrounding nozzle material, filament was prepared containing small amounts of tungsten powder as a contrast agent. During a third test, the extruder forces were measured and compared with the X-ray results and the predictions of a numerical simulation. The CT scans showed that at higher filament speeds, less area of the nozzle wall is in contact with the melt. This means that a larger part of the barrel section is occupied by an air gap between the solid filament and the nozzle wall. In contrast to the filament speed, the influence of the heater temperature shows no discernible effect on the part of the nozzle filled with melt. Radiographic evaluation of the velocity profile revealed a parabolic distribution under the studied conditions, closely matching numerical simulations modeling the flow as isothermal and non-Newtonian. The study's findings offer potential for improving nozzle design.  Furthermore, the presented experimental method can serve as a valuable tool for future validation of more complex numerical simulations.\end{abstract}

\clearpage
\section{Introduction}

\begin{figure}[h]
	\centering
		\includegraphics[scale=.54]{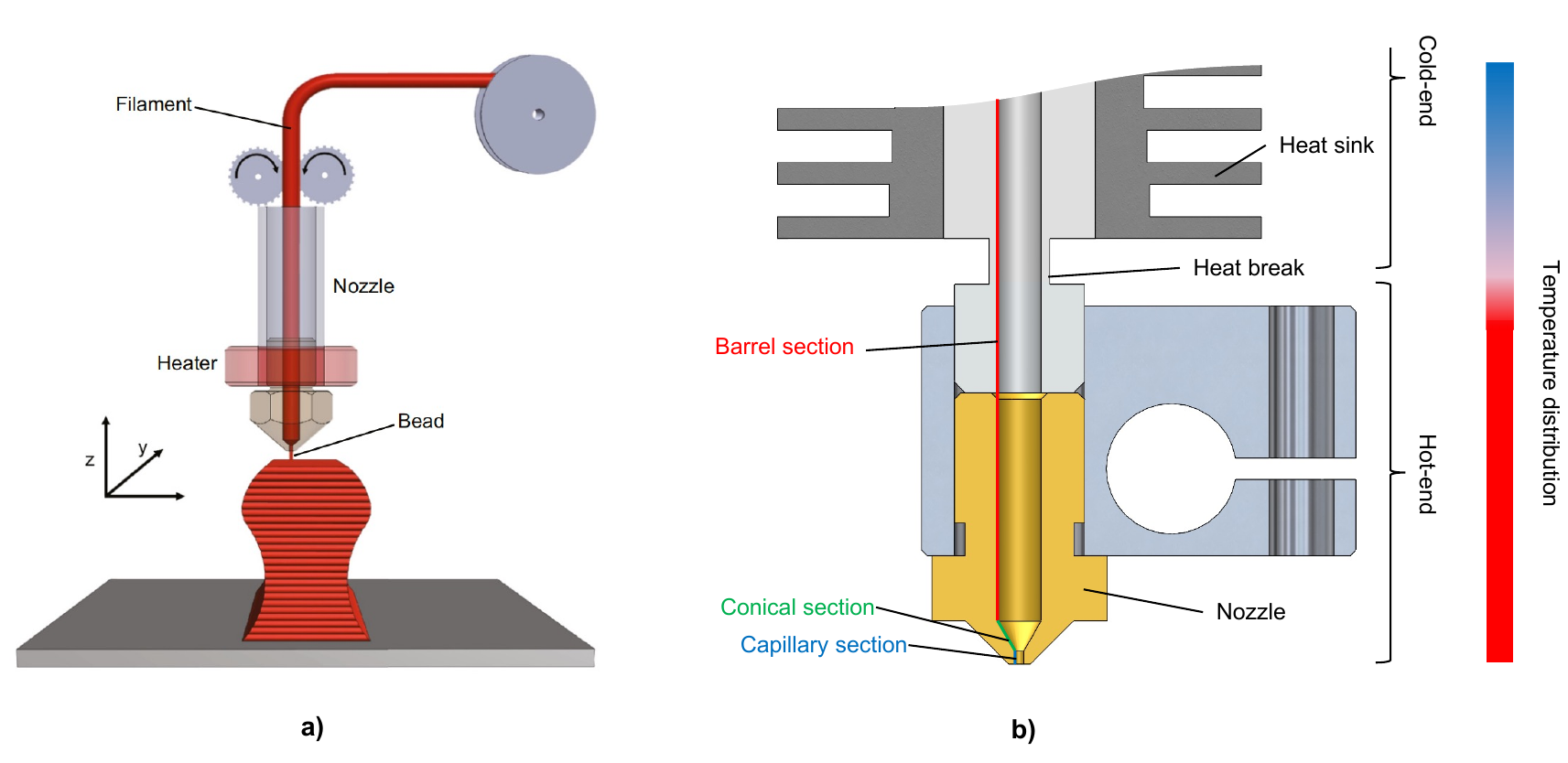}
	\caption{Schematic of (a) FFF process \cite{osswald_fused_2018}, and (b) the design of a typical printhead.}\protect
	\label{Fig1_SchematicFFF}
\end{figure}

 Due to the low initial cost of machine and material, Fused Filament Fabrication (FFF) is one of the most widely used additive manufacturing technique. Today, in addition to the rapidly growing hobbyist sector, the range of applications also includes many industrial sectors such as medicine or the aerospace industry, where FFF is used to produce complex and highly functional parts \cite{burkhardt_pandemic-driven_2020}. To produce an object, a pinch roller mechanism feeds a plastic filament into a print head mounted on a gantry. In the print head, the filament is molten and forced through a nozzle to extrude a thin strand. As the gantry moves the print head in two dimensions (x-y plane), the strand is deposited on the build platform according to the print path. After completion of a layer, a new layer is deposited on top of the previous layer and fuses with it. As illustrated in Fig.~\ref{Fig1_SchematicFFF}(a), the layer-by-layer deposition is repeated along the z-axis until the part is completed.

The focus of this work is the printhead, whose main task is to enable stable extrusion with high melt rates at low pressure loss. These requirements result in the typical two-part structure of a printhead (Fig.~\ref{Fig1_SchematicFFF}(b)), consisting of the so-called hot-end and cold-end. The hot-end is comprised of the nozzle and a surrounding metal block in which a heating coil is embedded. While the hot-end heats and melts the filament, the adjacent cold-end is designed to hold the filament in the solid phase to provide low surface friction with the tube through which the filament is fed. To achieve this, the cold-end consists of a heat sink cooled by a fan. In addition, to minimize heat transfer between the zones, a thermal break is used as a connecting element.\par
Various experiments have shown that filament melting and material extrusion can be limiting factors for the overall build rate \cite{go_fast_2017}. In addition, material extrusion problems are a major cause of failure during printing \cite{taheri_thermal_2022}. Because of this importance, many efforts have been made to gain a better understanding of the melting and extrusion behavior through analytical models, numerical simulations, and experiments.\par 

\begin{figure}[h]
	\centering
		\includegraphics[scale=.54]{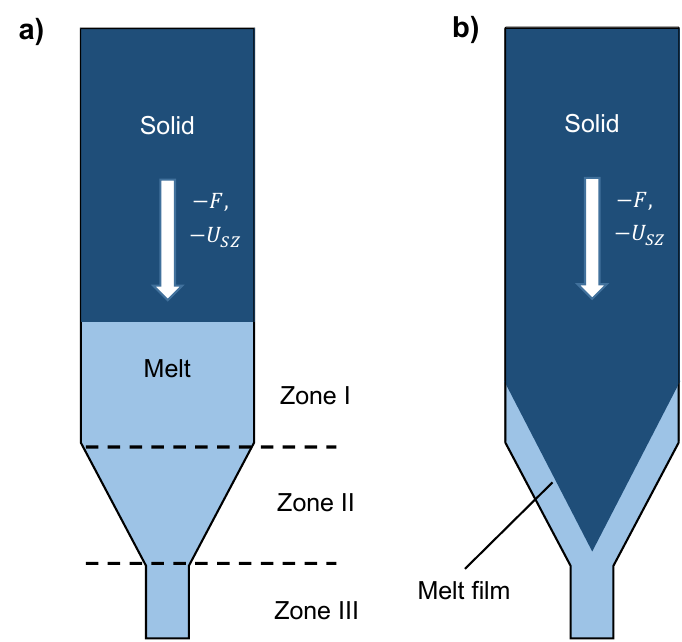}
	\caption{Schematic of (a) Bellini’s assumption \cite{bellini_liquefier_2004} and (b) Osswald’s assumption \cite{osswald_fused_2018} of the melting mechanism within the nozzle.}\protect
	\label{Fig2_Bell_Oss}
\end{figure}

There are two primary analytical models that attempt to explain how the melt flows in a printing nozzle. The first of these models postulates the existence of an isothermal pressure flow within the nozzle, which is generated by the solid filament. This model is referred to as the low speed model \cite{oehlmann_modeling_2021} and was initially proposed by Bellini et al. \cite{bellini_liquefier_2004}. Fig.~\ref{Fig2_Bell_Oss}(a) illustrates the low speed model, which entails the computation of the pressure drop in the nozzle as the sum of the pressure drop in three different sections: the barrel, the conical section and the capillary.\par

Another model, proposed by Osswald et al. \cite{osswald_fused_2018} and later improved by Colón et al. \cite{colon_quintana_implementation_2020}, represents a high speed scenario (Fig.~\ref{Fig2_Bell_Oss}(b)). It is assumed that the filament does not melt until it reaches the conical section of the nozzle, where a thin melt film forms. To determine the extrusion force, a mass, momentum and energy balance within the melt film is solved, while the Stefan boundary condition is used to calculate the thickness of the melt film.\par 

Due to the lack of experimental investigation of melting and flow behavior, it is still unclear under which conditions each model applies with one type of melting transition or the other. However, it is likely that for a wide range of filament speeds, the melting mechanism lies between the low and high speed scenarios, as suggested by numerical simulations \cite{kattinger_numerical_2022}.\par

The main differences in the published numerical models, apart from the viscosity model used, are the boundary conditions applied in the barrel section. In \cite{mostafa_study_2009} and \cite{ramanath_melt_2008} throughout the flow section a constant wall temperature equal to the heater temperature was assumed together with a no slip boundary condition. Under these conditions the filament melts shortly after entering the nozzle, resulting in isothermal flow over large parts of the nozzle. Another class of models assumes wall slip and radiative heat transfer mechanism for the barrel section or portions of it to account for an air gap between the filament and the wall \cite{kattinger_numerical_2022, phan_computational_2020}. However, these models require an a priori assumption regarding the expansion of the air gap in the axial direction. One result of these models is a strong non-isothermal flow in the capillary section and even at the outlet at higher filament speeds, which was also found in experiments. The numerical model of Serdeczny et al. differs from the previously mentioned models in that the volume of fluid (VOF) method is used to account for the free surface area between the incoming filament and the barrel wall \cite{serdeczny_experimental_2020}. A main prediction is the presence of recirculation region immediately after filament inlet, followed downstream by an isothermal flow. Based on the published numerical models, it is obvious that the question of how the gap between the incoming filament and the nozzle is filled is still unresolved.\par
In order to better understand the material extrusion and to validate the results of models and simulations, several experimental studies have been carried out. These include the use of force sensors \cite{mazzei_capote_trends_2021} or the measurement of the extrusion force with a load cell \cite{go_fast_2017, serdeczny_experimental_2020, kattinger_numerical_2022}. In addition, infrared cameras have been used in \cite{serdeczny_experimental_2020} and \cite{phan_rheological_2018} to measure the strand temperature directly below the nozzle exit. The above mentioned experimental studies have in common that they allow indirect conclusions to be drawn about the flow behavior. This distinguishes them from the experimental method of Peng et al., which allows direct observation of the flow behavior by introducing dye markers into a filament \cite{peng_complex_2018}. After extrusion of the prepared filament, the distribution of the dyes within the strand and inside the nozzle was visualized by optical microscopy and computed tomography respectively. For the latter study, the extrusion had to be stopped, the nozzle disassembled and then cut open to remove and characterize the solidified plastic. The evaluation of the pigment distribution in the nozzle at two different axial positions allowed conclusions to be drawn about the velocity profile. One noticeable result was a clear deviation from the ideal profile of a pressure-driven isothermal flow. However, since extrusion must be stopped in this method, effects such as gap filling cannot be investigated. This was made possible by the in-situ observation presented by Hong et al. using a glass nozzle and different colored filaments \cite{hong_-situ_2022}. One limitation is that the results obtained with a glass nozzle cannot be directly transferred to the typically used metal nozzles due to non-comparable material properties such as heat transfer or surface friction.\par

This paper presents an experimental method that allows in-situ observation of the melt flow in a metal hot-end. This is achieved by X-ray micro-computed tomography using a specially developed setup originally described in \cite{ehrler_ct-analysis_2023}. For this purpose, filament was prepared containing small amounts of tungsten powder as a contrast agent. Two types of experiments were performed with the prepared filament. In one experiment, a commercial hot-end was used, and the focus was on the visualization of the gap between the filament and the nozzle. In a second experiment, a nozzle optimized in terms of contrast (X-ray absorption) was used to study the velocity distribution in the nozzle and to provide a basis for validation of models and simulations. The experiments are complemented by measurements of the required extrusion force using a separate experimental setup. Based on these findings, a comparative analysis is presented, discussing the agreement between our experimental results and a simple numerical model.

\section{Materials and methods}
\subsection{Materials}

The impact modified polystyrene PS486N from INEOS Styrolution (Frankfurt am Main, Germany) was chosen for this work. Tungsten powder (Werth-Metall, Erfurt, Germany) with an average particle size of \SI{<6.3}{\um} was used as a contrast agent for the filament. This powder material was chosen because of its high atomic number ($_{74}^{}\textrm{W}$) and density (\SI[per-mode = symbol]{19.25}{\g\per\cm\cubed}), which according to the Lambert-Beer law are the decisive factors for high attenuation of X-rays. A volume fraction of 1\,\% (16.2 wt\%) was chosen for the experiments as a compromise between achieving good contrast with the surrounding nozzle and minimizing the impact on the rheological and thermal melt properties. To evaluate the latter aspect, a  comparison of the rheological and thermal properties was conducted between the filled and unfilled materials.

\subsection{Compounding and filament extrusion }
To compound the tungsten powder into the plastic melt, a twin-screw extruder equipped with a gravimetric feeder was employed. Subsequently, the extrudate obtained was ground into granules, which were further processed using a single-screw extruder to produce filaments with a diameter of \SI{1.75}{\mm}. The application of a laser micrometer allowed for monitoring of the filament diameter. A high-resolution CT scan validated the homogeneous distribution of tungsten particles across the filament cross section. More information on the evaluation can be found in the Appendix (\ref{Particle_distribution}).

\subsection{Material properties}
\subsubsection{Rheological characterization}
The rheological properties of the produced filament were measured using a TA Instruments Discovery HR-3 rheometer (TA Instruments, Inc., New Castle, USA) with a \SI{25}{\mm} parallel plate geometry and a gap of \SI{1}{\mm}. The dynamic viscoelastic properties were determined in the frequency range of 0.01–628 \si[per-mode = symbol]{\radian\per\s} at a strain value of 5\,\%, which was confirmed to be within the linear viscoelastic range. The material was tested at temperatures of \SI{205}{\celsius}, \SI{220}{\celsius}, and \SI{235}{\celsius}. By applying the Cox-Merz rule, the complex viscosity was converted to the steady-state shear viscosity \cite{cox_correlation_1958}.

\subsubsection{Differential scanning calorimetry (DSC)}
The temperature-dependent specific heat was determined using a Mettler Toledo DSC 2 (Mettler-Toledo, LLC, Columbus, USA). In accordance with ISO 11357, the following three measurements were conducted: (1) measurement of two empty pans to establish the baseline, (2) calibration measurement using sapphire, and (3) measurement of the sample. All measurements were carried out under identical conditions.\par
The experiment began at an initial temperature of \SI{20}{\celsius}, and the temperature was then gradually increased to \SI{250}{\celsius} at a rate of \SI[per-mode=symbol]{10}{\K\per\minute}. This was followed by an isothermal phase lasting \SI{5}{\minute}, after which a second heating was performed, again increasing the temperature to \SI{250}{\celsius} at the same rate. To account for the influence of previous processing, the heat capacity was determined based on the first heating curve.

\subsection{Experimental setups}

\begin{figure}[h]
	\centering
		\includegraphics[scale=0.5]{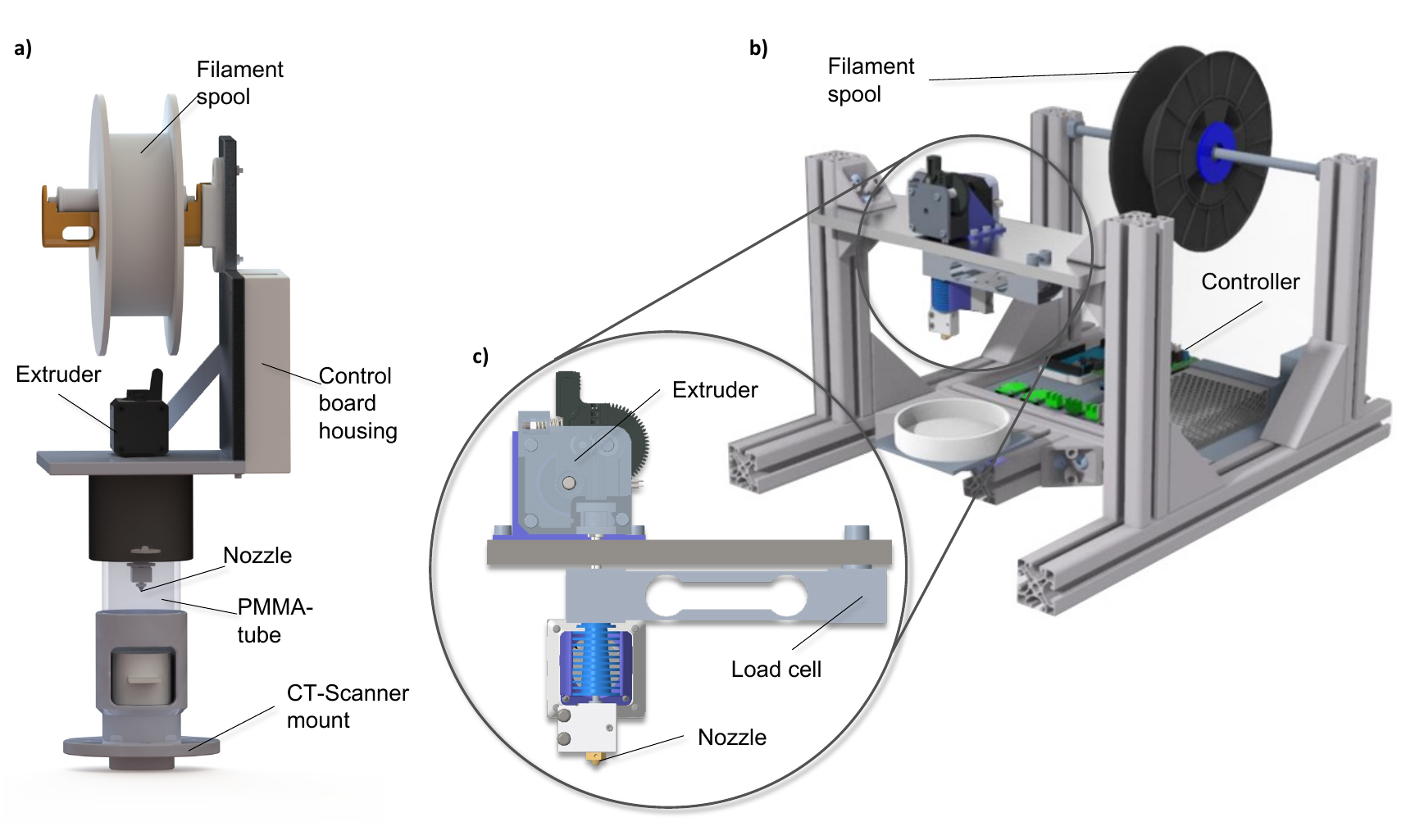}
	\caption{Schematic of (a) the in-situ CT scanning setup and (b) the extrusion force measurement setup with (c) a detailed view of the load cell used.}\protect
	\label{Fig4_ExperimentalStands}
\end{figure}

For this work, the experimental setup shown in Fig.~\ref{Fig4_ExperimentalStands} was developed, which can be mounted and operated inside a CT scanner. The experimental setup is based on a commercial extruder (model LGX form Bondtech, Värnamo, Sweden) that was equipped with a E3D-V6 type hot-end. A PC connected to a Duet 2 Ethernet board (Duet3D, Peterborough, United Kingdom) with the RepRap firmware installed was used to control the setup. To enable an X-ray scan with high resolution, the experimental setup was designed to be rotated at the smallest possible distance from the X-ray source during extrusion. Care was taken in the design to ensure that the region of interest remains unobscured by metal parts during scanning. This involved routing cables through a top channel and utilizing a PMMA tube as a connecting element between the lower and upper sections of the setup.\par
A second experimental setup was used to measure the extrusion force required to push the filament through the nozzle. The measurement principle is based on a load cell mounted between the extrusion mechanism and the hot-end. As shown in Fig.~\ref{Fig4_ExperimentalStands}(c), the heat sink is screwed directly into the load cell through which the filament passes. In this way, the axial force exerted by the feeding mechanism on the filament is measured. For controlling the setup, a PC connected to an Arduino Mega 2560 microcontroller and a RAMPS 1.4 board was used. A second Arduino Mega 2560 was used for data acquisition. For more details about the setup used, please refer to \cite{kattinger_numerical_2022}.                                  

\subsection{In-situ micro computed tomography}

The experimental setup for X-ray observation was mounted in a YXLON FF20 ${\mu}$-CT Scanner equipped with a FXE transmission beam X-ray tube (YXLON International GmbH, Hamburg, Germany) and a Varex 2530HE (Varex Imaging, Salt Lake City, Utah, USA) detector providing 16-bit image depth. In this work, the ${\mu}$-CT Scanner was used to generate both 360°-CT data and 2D data by projectional radiography. For both types of experiments, a tube voltage of \SI{190}{\kV} and a current of \SI{60}{\uA} were used, corresponding to an X-ray system power of \SI{11.4}{\W}. These settings required the use of a high-performance transmission target. To reduce noise and shorten the measurement time, 2x2-pixel binning, and a sensitivity of \SI{2.00}{\pF} were used. 

\subsubsection{CT scan}

In one experiment, a 360°-CT scans were performed to investigate how the melting behavior of the filament depends on the filament speed and heater temperature by examining the air gap between the incoming filament and the nozzle wall. For the experiments a commercially available titanium nozzle with a capillary diameter of 0.4 mm was used. The nozzle details are shown in Fig.~\ref{Fig5_Nozzle_CT_Radiation} and Table~\ref{Nozzle geometry parameters}. Three different heater temperatures (\SI{220}{\celsius}, \SI{240}{\celsius}, \SI{260}{\celsius}) and filament speeds (\SI[per-mode = symbol]{0.5}{\mm\per\s}, \SI[per-mode = symbol]{2.0}{\mm\per\s}, \SI[per-mode = symbol]{4.0}{\mm\per\s}) were tested. These values were selected based on the recommendations provided by a manufacturer of comparable high impact modified polystyrene and cover a typical printing speed range \cite{noauthor_basf_nodate}.\par
After setting the specified filament speed, the material was extruded for \SI{30}{\s} to ensure that quasi-stationary behavior was achieved. The CT scan was then started, which consisted of 720 images and required \SI{270}{\s} to complete. For reconstruction, the Comet Yxlon FF CT software was used, which is based on the Siemens CERA software package (Siemens Healthcare GmbH, Erlangen, Germany) and employs an FDK algorithm \cite{feldkamp1984practical}. To reduce computational time and data volume, only the region of interest was covered for reconstruction, resulting in a data field of 586 x 1072 x 880 voxels with a voxel size of \SI{27.6}{\um}. The CT data analysis was performed with the software Avizo 2020.3 (Thermo Fisher Scientific Inc., Waltham, Massachusetts, USA).

\subsubsection{Observation of the flow behavior}

\begin{figure}[h]
	\centering
		\includegraphics[scale=0.5]{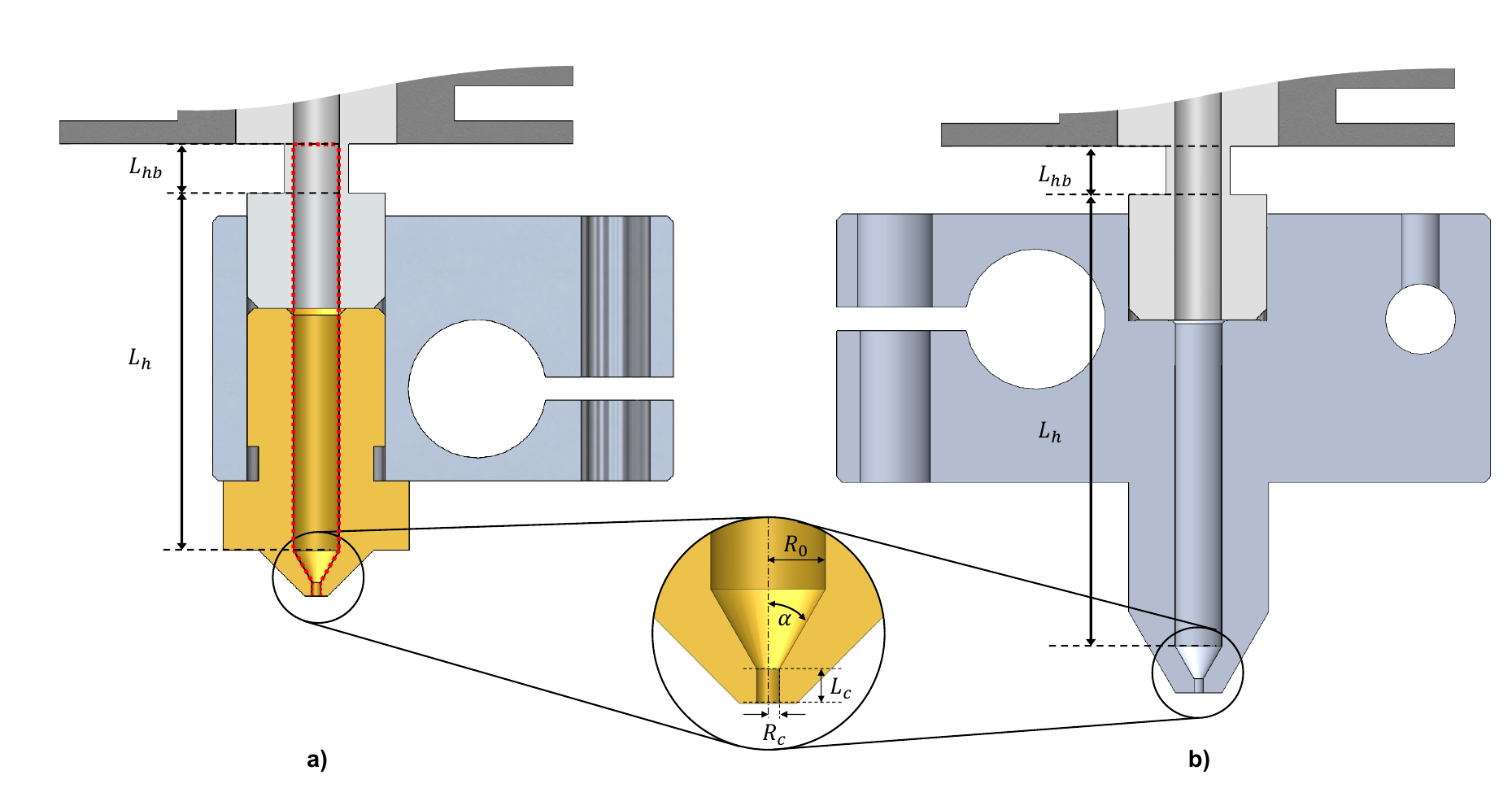}
	\caption{Schematic of the hot-end used for (a) 360° CT scans and for (b) the radiography. For the 360° CT scans, the area examined is highlighted by the red dashed lines.}\protect
	\label{Fig5_Nozzle_CT_Radiation}
\end{figure}

In another experiment, the flow behavior in the nozzle was observed by video recording using projectional radiography. Due to the high X-ray attenuation, the portion of the nozzle screwed into the steel heating block would be difficult to examine in this way. Therefore, as shown in Fig.~\ref{Fig5_Nozzle_CT_Radiation}(b), an aluminum heating block optimized with respect to X-ray attenuation was manufactured, in which the nozzle is directly integrated.\par
\begin{figure}[h]
	\centering
		\includegraphics[scale=0.5]{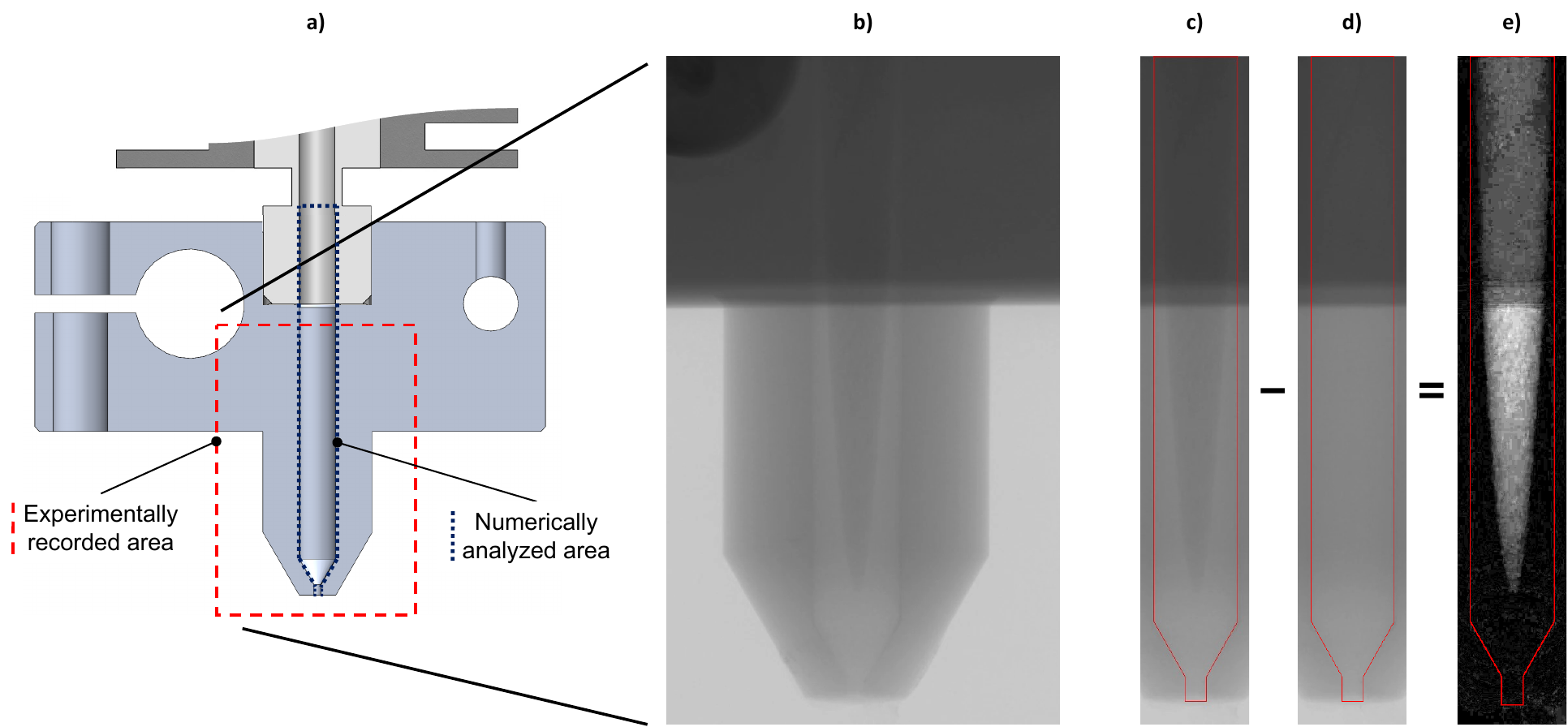}
	\caption{Schematic of the hot end used for radiography (a). The red dashed lines mark the recorded area. Image (b) shows the resulting X-ray image of this area. Also included are two images cropped to the region of interest (ROI). Image (c) shows the barrel section partially filled with the higher absorbing melt, while image (d) is the background image where the nozzle is filled only with pure polystyrene. Image (e) is the difference image obtained by subtracting image (d) from image (c). The difference image exhibits a contrast difference that is due to a wall thickness jump in the upper part of the examined nozzle.}\protect
	\label{Figx2_Nozzle_CT_Radiation}
\end{figure}

To enable the observation of the flow behavior, the nozzle was first filled with pure polystyrene. The filament containing the tungsten powder was then fed into the nozzle via the feeding mechanism at a defined speed while X-ray images were taken at a frequency of \SI{2}{\Hz}. This was done at a heater temperature of \SI{240}{\celsius} for three different filament speeds: \SI[per-mode=symbol]{0.5}{\mm\per\s}, \SI[per-mode=symbol]{1.0}{\mm\per\s}, and \SI[per-mode=symbol]{2.0}{\mm\per\s}. As soon as the pure polymer melt was completely replaced by the higher absorbing melt, the recording was stopped. Due to the limitations of the system, exporting the video or individual images was not possible. Therefore, the software OBS was utilized to capture the screen output of the live view generated by the CT scanner. This resulted in an 8-bit video, from which the necessary images were extracted. Fig.~\ref{Figx2_Nozzle_CT_Radiation} highlights the investigated portion of the nozzle and shows an exemplary X-ray image. The image processing involved cropping the images to fit the region of interest (ROI) and subtracting the background corresponding to the nozzle filled with pure melt to obtain a difference image. All image analysis and processing were performed using MATLAB (MathWorks, Natick, Massachusetts, USA).

\begin{table}[H]
\centering
\caption{Nozzle geometry parameters}
\label{Nozzle geometry parameters}
\begin{tabular}{ccccccc}
\hline
Nozzle Type     & \begin{tabular}[c]{@{}c@{}}$R_c$\\ (mm)\end{tabular} & \begin{tabular}[c]{@{}c@{}}$R_0$\\ (mm)\end{tabular} & \begin{tabular}[c]{@{}c@{}}$L_c$\\ (mm)\end{tabular} & \begin{tabular}[c]{@{}c@{}}$L_{hb}$\\ (mm)\end{tabular} & \begin{tabular}[c]{@{}c@{}}$\alpha$\\ (deg)\end{tabular} & \begin{tabular}[c]{@{}c@{}}$L_h$\\ (mm)\end{tabular} \\ \hline
CT scan (a)     & \multirow{2}{*}{0.2}                                 & \multirow{2}{*}{1.0}                                   & \multirow{2}{*}{0.6}                                 & \multirow{2}{*}{5.0}                                    & \multirow{2}{*}{30}                                      & 10.51                                                \\ \cline{1-1} \cline{7-7} 
Radiography (b) &                                                      &                                                      &                                                      &                                                         &                                                          & 20.50                                                      \\ \hline
\end{tabular}
\end{table}

\subsection{Extrusion force measurement}
The measurements were carried out  with the in-house manufactured nozzle. Before the test, a calibration of the feeding mechanism ensured that the commanded filament speed matched the actual filament speed. The experimental procedure consists of measuring the extrusion force at a frequency of \SI{1}{\Hz} for \SI{240}{\s} while the filament gets extruded at a defined speed. To consider only the steady state, the first 60 s of the measurement was excluded for the subsequent calculation of the mean and standard deviation. Collecting the extruded strand directly under the nozzle ensured that the weight of the strand had a negligible influence on the result. 

\subsection{Numerical modeling}
As part of this work, numerical simulations were performed assuming isothermal conditions and wall adhesion throughout the nozzle. The assumption of isothermal flow simplifies reality significantly by disregarding the melting process and the transition between the solid filament and the melt flow. However, previous studies suggest that the isothermal flow assumption may be a reasonable approximation within the range of filament speeds investigated here \cite{kattinger_numerical_2022, serdeczny_numerical_2020}. The objective of this study is to evaluate this by comparing the experimental results with the numerical simulation. This analysis aims to provide deeper insights into the thermal behavior and wall adhesion effects in the nozzle, thus contributing to a better understanding of the filament melting.\par
\begin{figure}[h]
	\centering
		\includegraphics[scale=0.5]{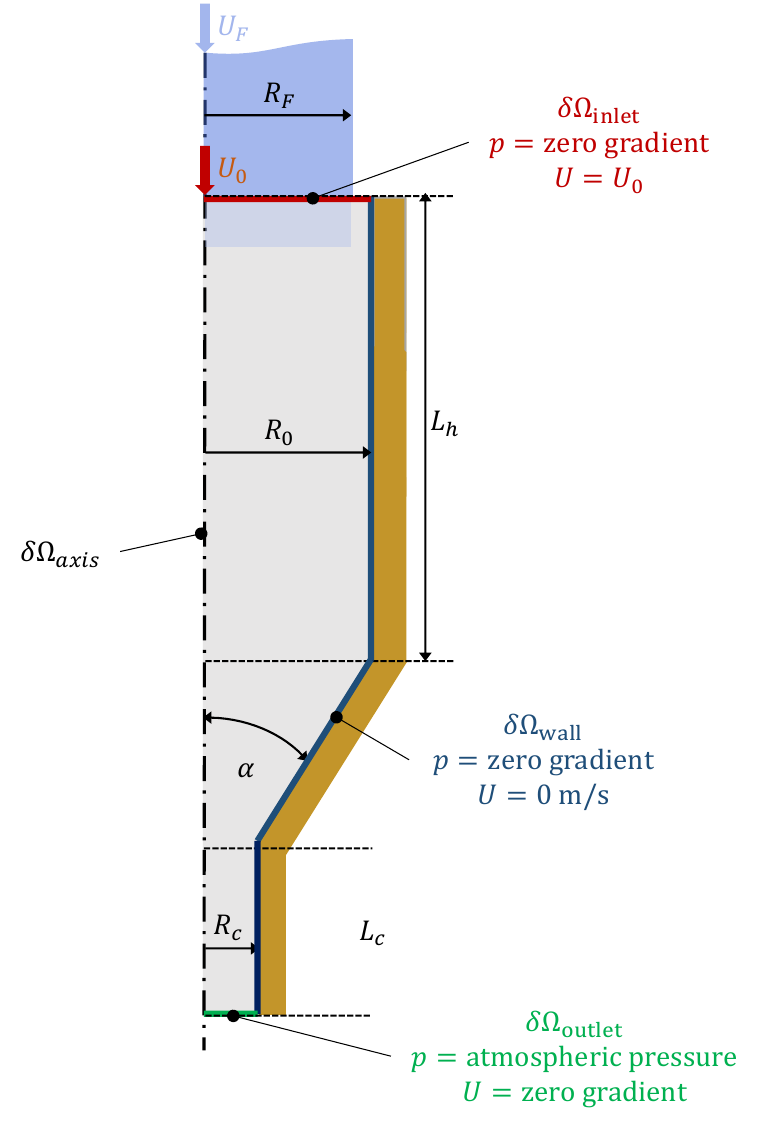}
	\caption{Simulation model and boundary conditions.}\protect
	\label{Fig6_boundary_condition}
\end{figure}

Fig.~\ref{Figx2_Nozzle_CT_Radiation} highlights the specific part of the hot end considered for the simulations, encompassing all parts downstream of the thermal break and spanning a barrel length denoted as $L_h$ (see Fig.~\ref{Fig5_Nozzle_CT_Radiation} and Tab.~\ref{Nozzle geometry parameters}). In Fig.~\ref{Fig6_boundary_condition}, a schematic representation of the simulation model is shown, illustrating the applied boundary conditions. To ensure consistency with the experimental setup, a constant velocity was assigned as the inlet boundary condition. The relationship between the experimentally selected filament speed $U_F$ and the applied inlet boundary condition $U_0$ can be expressed by Eq.~\ref{eq: Equ_1}.

\begin{equation} \label{eq: Equ_1}
		\begin{aligned}
        \frac{U_F}{U_0}=\frac{{R_0}^2}{{R_F}^2} 
        \end{aligned}
\end{equation}

The simulations were performed in steady-state mode using the open-source CFD package OpenFOAM v10 \cite{the_openfoam_foundation_openfoam_2011}. To simplify the problem, an axisymmetric model of the nozzle was employed, along with a structured 2D mesh comprising 25 elements over the capillary radius. The governing equations for the polymer flow, under the specified assumptions, are as follows:

\begin{equation} \label{eq: Equ_2}
	\nabla \cdot (\rho \underline{u} )=0
\end{equation}

\begin{equation} \label{eq: Equ_3}
	\nabla \cdot (\rho \underline{u} \underline{u} )=\nabla \cdot \underline{\underline{\tau } } -\nabla p
\end{equation}

\noindent Here, $\rho$ represents the density of the polymer, $\underline{u}$ is the velocity vector, $\underline{\underline{\tau}}$ is the deviatoric stress tensor, and $p$ is the pressure. To model the  melt flow, a Generalized Newtonian Fluid approach is used, where the deviatoric stress tensor is defined as:

\begin{equation} \label{eq: Equ_4}
	\underline{\underline{\tau } }={\eta}[\nabla \underline{u} + (\nabla \underline{u})^{T}]={\eta}\underline{\underline{S } }
\end{equation}

\noindent Here, $\underline{\underline{S}}$ represents the strain rate tensor, and $\eta$ is the dynamic viscosity. To account for non-Newtonian effects, the viscosity is modeled using the Carreau-Yasuda model:

\begin{equation} \label{eq: Equ_5}
	\frac{\eta}{\eta_0}=\left(1+\left(\lambda\dot{\gamma}\right)^a\right)^\frac{n-1}{a}
\end{equation}

\noindent In Eq.~\ref{eq: Equ_5}, $\eta_0$ is the zero shear viscosity, $\lambda$ is the relaxation time, $n$ is the power index, $a$ is the transition index, and $\dot{\gamma}$ is the magnitude of the strain rate tensor defined as:

\begin{equation} \label{eq: Equ_6}
	\dot{\gamma}=\sqrt{\frac{1}{2}\underline{\underline{S }}:\underline{\underline{S }}}
\end{equation}

To allow comparison with experimental results of the phase boundary between the filled and unfilled melt, timelines were calculated using the software Paraview \cite{ahrens200536} in a post-processing step. The streamline integration necessary for this calculation was performed utilizing the 4th Order Runge-Kutta method.
\section{Experimental results and Discussion}

\subsection{Material properties}
The viscosity data of the tungsten powder-filled and unfilled material (Appendix \ref{Rheometric data}) was used to create a master curve at \SI{240}{\celsius} using TA Instruments' Trios software (TA Instruments, USA). To determine the shift factor in the x-direction, the Time-Temperature Superposition (TTS) principle based on the Williams-Landel-Ferry (WLF) model was employed. A reference temperature of \SI{220}{\celsius} was selected for the shift.

\begin{equation} \label{eq: Equ_7}
	loga_{WLF}=\frac{{-C}_1\left(T-T_{ref}\right)}{C_2+\left(T-T_{ref}\right)}
\end{equation}

\begin{figure}[h]
	\centering
		\includegraphics[scale=1]{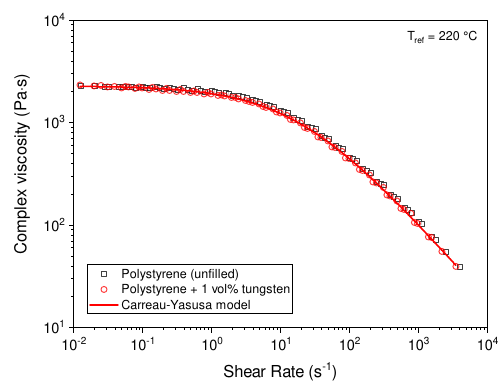}
	\caption{Master curve of shear rate versus complex viscosity for unfilled and tungsten powder filled polystyrene.}\protect
	\label{Fig3_Fig7_Master_curve}
\end{figure}

Fig.~\ref{Fig3_Fig7_Master_curve} shows the two master curves obtained and provides a comparison showing a minimal influence of powder addition on the rheological properties. The parameters of the Carreau-Yasuda model were determined using the data from the tungsten powder-filled material. The resulting values for the WLF model parameters are $C_1 = 8.068$ and $C_2 = \SI{171.514}{\K}$. Fitting the Carreau-Yasuda model (Eq. \ref{eq: Equ_6}) yields the values \SI{2300}{\Pa\s}, \SI{0.054}{\s}, 0.24, and 0.63 for $\eta_0$, $\lambda$, $n$, and $a$, respectively.

\begin{figure}[h]
	\centering
		\includegraphics[scale=1]{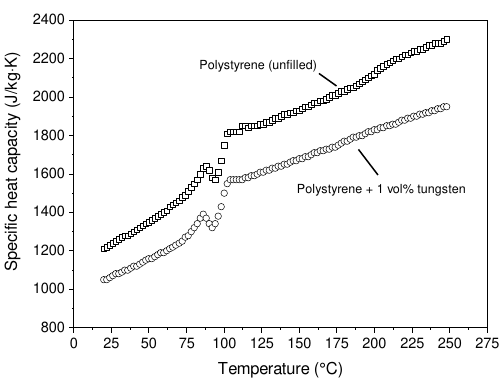}
	\caption{Temperature dependent specific heat capacity of the unfilled and tungsten powder filled polystyrene measured by DSC.}\protect
	\label{Fig3_cp}
\end{figure}

To evaluate the effect of tungsten addition on thermal properties, the specific heat capacity of the filled material was measured as a function of temperatures and compared with that of the unfilled material. The specific heat capacity data for the unfilled material was previously published in \cite{kattinger_numerical_2022}. Fig.~\ref{Fig3_cp} shows the comparison, indicating a noticeable shift on the y-axis due to the inclusion of tungsten powder. Interestingly, the glass transition temperature ($T_g$) remained unaffected by the tungsten powder addition, with $T_g$ measuring \SI{101}{\celsius} for both the filled and unfilled polystyrene. This finding suggests that the results obtained in this study using the metal powder-filled filament are also representative of unfilled filaments.

\subsection{Results obtained through CT scans}

\begin{figure}[h]
	\centering
		\includegraphics[scale=0.5]{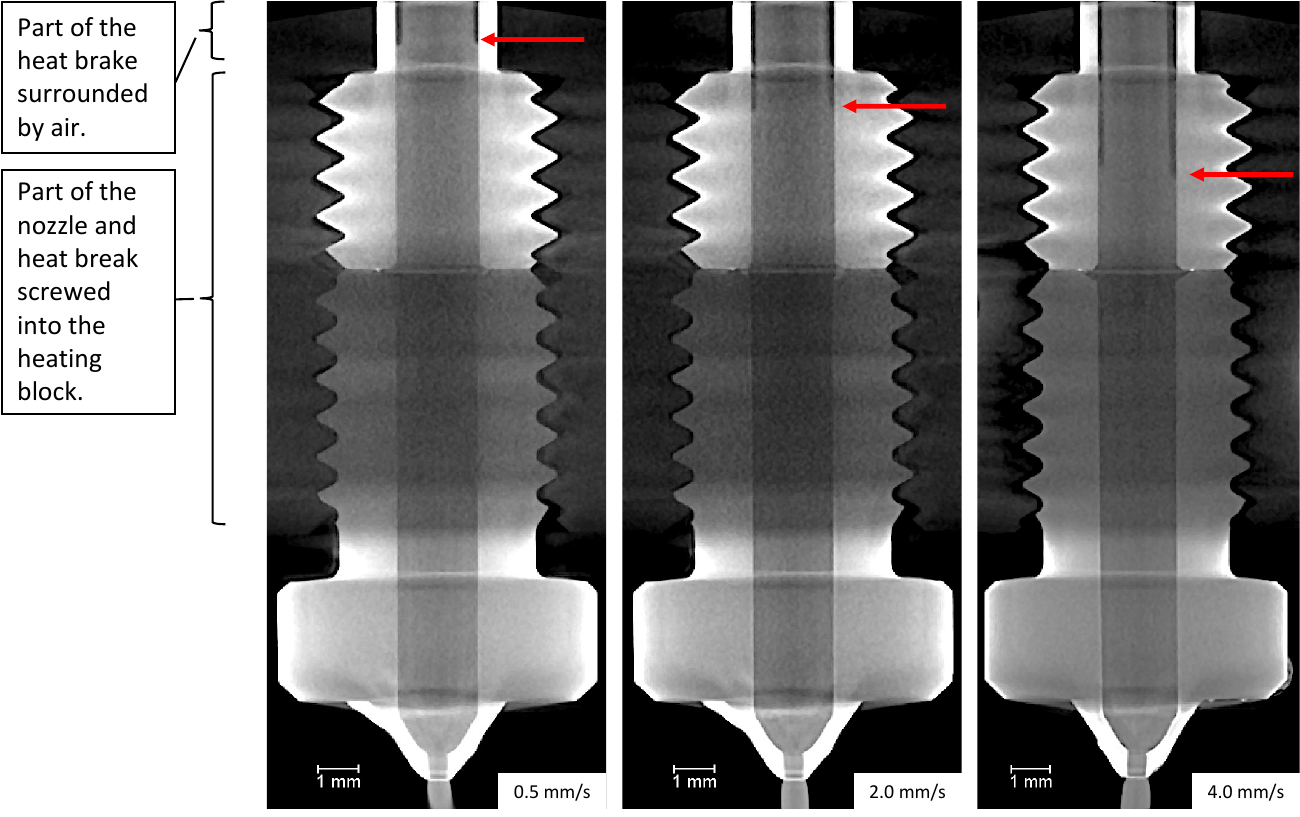}
	\caption{Cross-sectional images of CT scans generated at different filament speeds.}\protect
	\label{Fig9_CT_cross_section}
\end{figure}

\begin{figure}[h]
	\centering
		\includegraphics[scale=0.55]{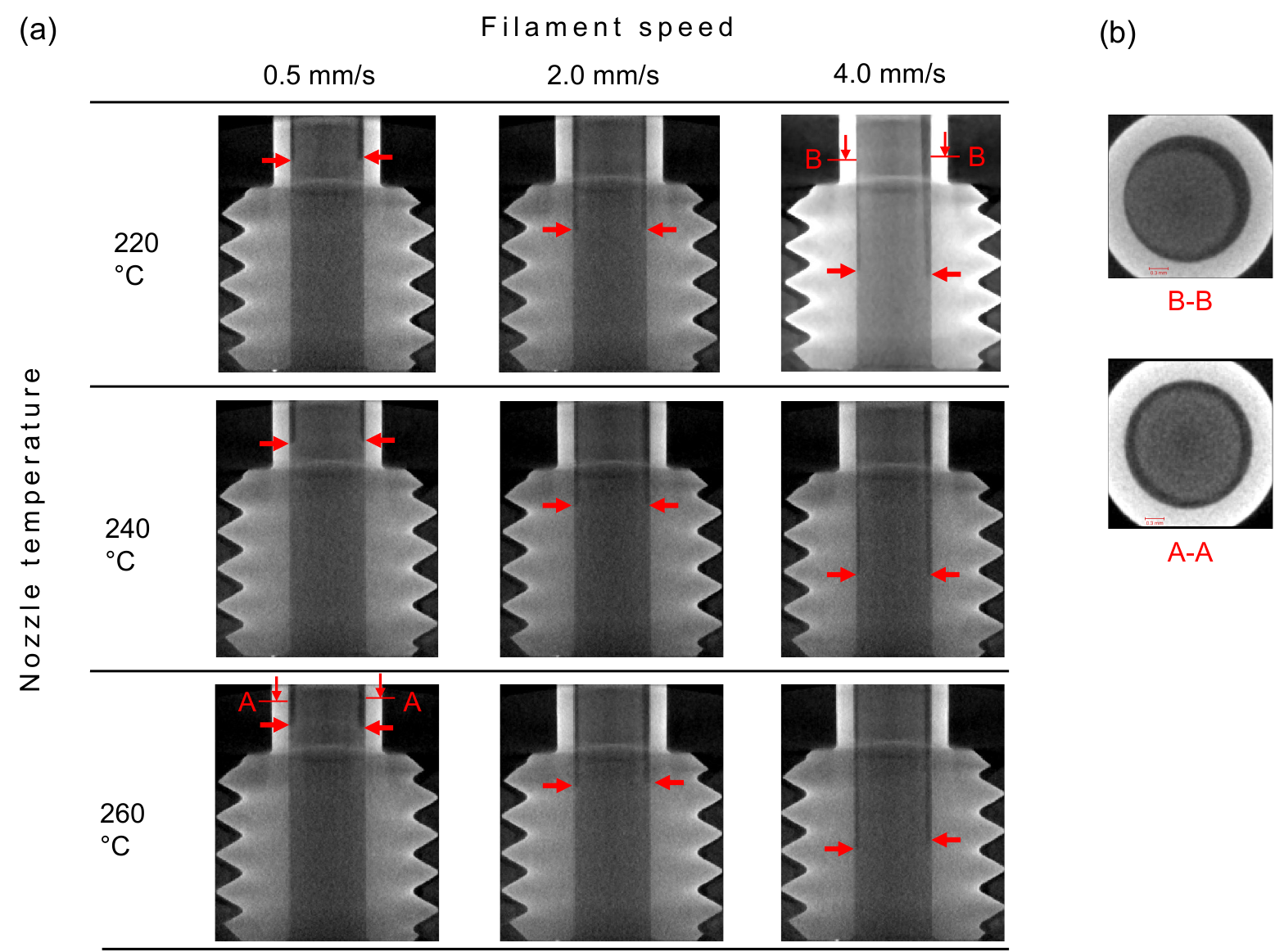}
	\caption{(a) Detailed view of gap filling as a function of heater temperature and filament speed, (b) Cross-sectional view showing the influence of filament speed.}\protect
	\label{Fig7_Deataild_gap_filling}
\end{figure}

Fig.~\ref{Fig9_CT_cross_section} shows three cross-sectional images of CT scans, each generated at a different filament speed. Based on the CT scans, an edge can be seen between the polymer melt and the surrounding air despite the scan duration of \SI{270}{\s}. This indicates that steady state conditions prevail. Otherwise, blurred edges would be expected in a CT scan. Furthermore, the results show an influence of the filament speed on the extent to which the melt is in contact with the walls in the barrel section. A comparison of all three images shows that the contact area between the melt and the wall decreases with increasing filament speed. This is due to the fact that the higher the filament speed, the less time the filament has to melt when passing through the nozzle. As shown in Fig.~\ref{Fig7_Deataild_gap_filling}(a), the heater temperature has no discernible effect on the contact area between the melt and the wall in the barrel section.  An interesting fact from the experimental results is that at the lowest filament speed selected, the gap fills up to the heat break. It is assumed that this is the maximum possible gap filling height, due to the sharp temperature drop in the heat break. Another aspect that was investigated using the CT data is the circularity of the air gap (Fig.~\ref{Fig7_Deataild_gap_filling}(b)). With increasing filament speed, a higher extrusion force is required to push the filament through the nozzle. This is accompanied by an increasing bending and buckling of the filament and thus decreasing concentricity of the air gap of the solid filament.\par

\subsection{Insights into the flow behavior}

\begin{figure}
	\centering
		\includegraphics[scale=1]{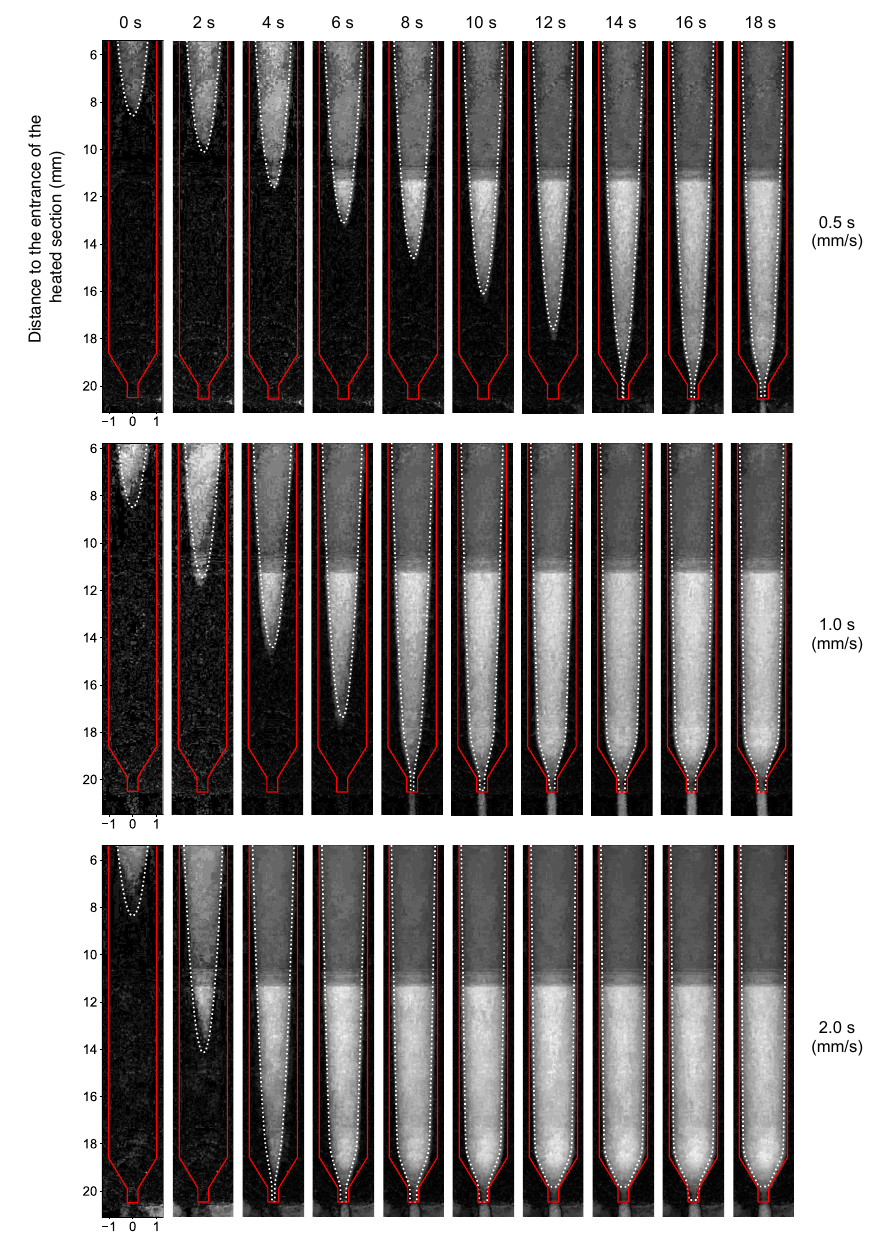}
	\caption{Results of the projectional radiography showing the influence of filament speed on the flow behavior. The dashed yellow curve corresponds to the numerical simulation.}\protect
	\label{Fig11_projectional_radiagraphy}
\end{figure}

Based on the second experiment, in which X-ray video recordings were made, the residence time distribution of the melt flowing in the nozzle can be studied. This in turn allows conclusions to be drawn about how the filament melts and how the flow in the nozzle develops. Fig.~\ref{Fig11_projectional_radiagraphy} shows frames of three different videos, each taken at a different filament speed, but the same heater temperature of \SI{240}{\celsius}. By post processing the images, the boundary between the pure polymer and the more absorbent tungsten powder loaded melt was made visible. 
In addition, Fig.~\ref{Fig11_projectional_radiagraphy} presents a comparison between the numerically solved time lines and the experimental results. The flow field solved numerically is provided in the appendix \ref{Flow field}. For the figure at initial time \SI[per-mode=symbol]{0}{\s}, a manually selected best matching pair of an experimental image and a numerically calculated time line was chosen. Subsequently, both the experimental images and the numerical solution were uniformly incremented by a time step of \SI[per-mode=symbol]{2}{\s} and compared.
The experimental results consistently show a parabolic interface between the pure and filled material at all three filament speeds studied, closely matching the corresponding numerical solution.  This implies that the assumed flow behavior, i.e., wall adhesion in the barrel section, accurately represents the real scenario.  Consequently, in the examined area, the filament has entered the molten state, with the air gap above the monitored region. These observations hold true throughout the range of filament speeds investigated, which encompass a slow to medium print speed range (approximately \SI[per-mode=symbol]{10}{\mm\per\s} to \SI[per-mode=symbol]{40}{\mm\per\s}). The investigation of higher filament speeds was constrained by the limited time resolution of the X-ray detector. 

\begin{figure}
	\centering
		\includegraphics[scale=0.9]{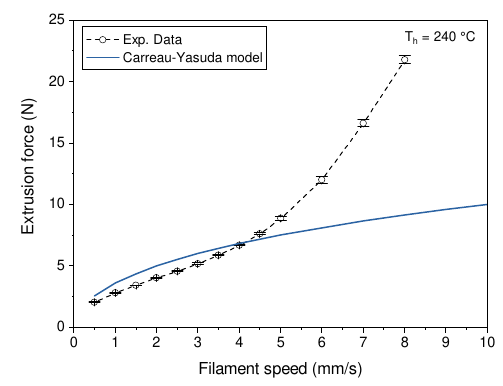}
     \caption{Comparison of the experimentally measured extrusion force with the predictions of the numerical simulation.}\protect 
	\label{ExperimentalSetup}
\end{figure}

Moreover, the comparison between the measured extrusion force and the numerical solution reveals a clear correlation: as depicted in Fig.~\ref{ExperimentalSetup}, the extrusion force consistently rises as filament speeds increase. Notably, the extrusion force measurements encompass a considerably broader range of filament speeds, including both the range examined via X-ray imaging and even higher speeds relevant to high-speed 3D printing. Up to a filament speed of \SI[per-mode=symbol]{3.0}{\mm\per\s}, the experimental curve exhibits an almost linear relationship between the extrusion force and the filament speed, with the numerical solution closely following this trend. However, with a further increase in filament speed, the trend changes to a steeper increase in the extrusion force. In this range, the simulation results clearly deviate from the experimental results, with the simulation showing significantly lower extrusion forces. This comparison of extrusion forces complements the insights obtained from the X-ray data and highlights the consistency of the results. Specifically, it demonstrates that in the range of low to moderate filament speeds (up to \SI[per-mode=symbol]{3.0}{\mm\per\s}), both the numerically solved time lines and the extrusion force closely align with the experimental findings. This agreement suggests that at these filament speeds, there is sufficient time for the filament to melt, aligning well with the assumption of isothermal flow. However, as the filament speed increases, the time available for complete melting of the filament becomes insufficient, leading to a deviation from isothermal flow. Consequently, the numerical simulations deviate from the experimental results in this range.

\section{Conclusion}
This paper presents an experimental method that allows in-situ observation of melt flow in a metal hot-end using X-ray micro-computed tomography. For this purpose, a filament containing small amounts of tungsten powder as a contrast agent was prepared. Two experiments were performed with the motivation to better understand how a filament melts and flows in the nozzle of an FFF printer. In one experiment, 360° CT scans were taken to observe how the wall contact area between the filament and the nozzle wall depends on the filament speed and the heater temperature. In a second experiment, radiography was used to study the flow profile that develops inside the nozzle at different filament speeds. The results obtained in this way were supplemented by measurements of the extrusion force and a comparison with a numerical simulation. Several conclusions can be drawn from the experiments:

\begin{itemize}
  \item The CT scans showed that at higher filament speeds, less area of the nozzle wall is in contact with the melt. This means that a larger part of the barrel section is occupied by an air gap between the solid filament and the nozzle wall.  In contrast to the filament speed, the influence of the heater temperature shows no discernible effect on the part of the nozzle filled with melt. CT scans can also be used to study phenomena such as filament buckling in the upper part of the print head. The results indicate that buckling increases at high filament speeds.
  \item An evaluation of the velocity profile revealed a parabolic velocity distribution with zero velocity at the wall and maximum velocity in the center for all filament speeds examined (\SI[per-mode = symbol]{0.5}{\mm\per\s}, \SI[per-mode = symbol]{1.0}{\mm\per\s}, \SI[per-mode = symbol]{2.0}{\mm\per\s}). The velocity profile has the shape that would be expected from a pressure flow.
  \item Based on the results obtained, the linear relationship observed between filament speed and extrusion force at lower filament speeds can be attributed to a sufficient melting time for the filament. This conclusion is supported by the strong agreement between the numerical solution of an isothermal flow and the X-ray data. The sharp increase in extrusion force at higher filament speeds, known as the transition zone, is recognized as a potential source of extrusion issues \cite{nienhaus_investigations_2019}. No such problems were detected within the range of tested filament speeds. Nevertheless, it is recommended, based on the findings, to operate the hot end under conditions that ensure sufficient melting time for the filament within the barrel section of the nozzle. If higher printing speeds are desired, this necessitates the use of a nozzle with an extended barrel section. 
\end{itemize}

In future research, higher temporal resolution, utilizing high-speed X-ray imaging, should be chosen to investigate higher filament speeds. This approach can be complemented by incorporating X-ray particle tracking velocimetry, as demonstrated in \cite{parker_enhanced_2022, makiharju_tomographic_2022}. Additionally, expanding the area of examination in the nozzle will allow for a more detailed analysis of phenomena like filament buckling or the transition of a solid plug flow to a parabolic melt flow. Achieving these goals would provide even more valuable results for the study of extrusion problems.\par
Future work should also include the investigation of other materials, such as semi-crystalline plastics. It is also worth investigating whether the proportion of contrast agent can be optimized for even better contrast. In addition, it is proposed to compare the experimental data with more advanced simulations, as this approach provides a novel method for validating the simulated flow behavior.

\section*{CRediT Authorship Contribution Statement}
\textbf{Julian Kattinger:} Writing, Data acquisition, Data analysis. \textbf{Mike Kornely:} Development of the test stand, Data acquisition, Data analysis. \textbf{Julian Ehrler:} Data acquisition, Data analysis, Writing. \textbf{Marc Kreutzbruck} and \textbf{Christian Bonten:} Supervision, Resources, Funding.

\section*{Declaration of Conflict of Interest}
The authors declare that there are no conflicts of interest, financial or personal, regarding the publication of this paper.

\section*{Acknowledgements}
This research was funded by the Deutsche Forschungsgemeinschaft (DFG, German Research Foundation) under grant number 503938087. 
\section{Appendix}

\subsection{Particle distribution}
\label{Particle_distribution}

Using a CT scanner, the particle distribution was examined. The scan involved X-ray settings of a tube voltage of \SI[per-mode = symbol]{70}{\kV} and a current of \SI{45}{\uA}. For each image, the exposure time was \SI[per-mode = symbol]{1.0}{\s}, and a total of 2610 images were acquired to complete a single CT scan. The resulting magnification produced a voxel size of \SI{1.4}{\um}.\par

\begin{figure}[H]
 \centering
 \begin{subfigure}[b]{.35\textwidth}
	\centering
		\includegraphics[width=\linewidth]{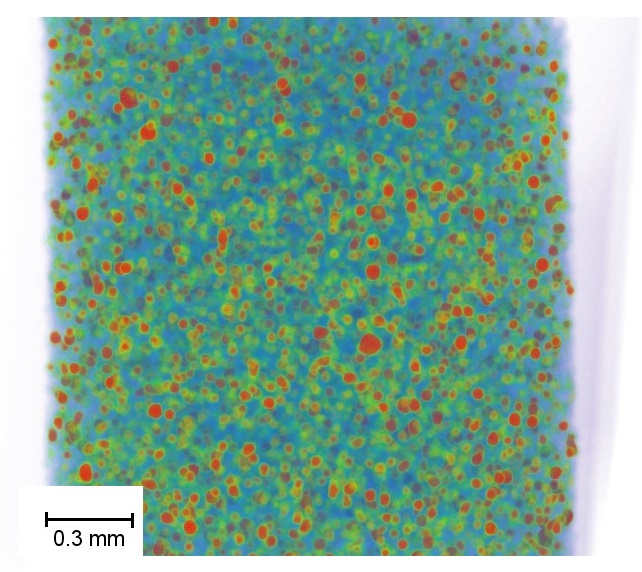}
	\caption{}
	\label{Particle1}
 \end{subfigure}
 \begin{subfigure}[b]{.35\textwidth}
	\centering
		\includegraphics[width=\linewidth]{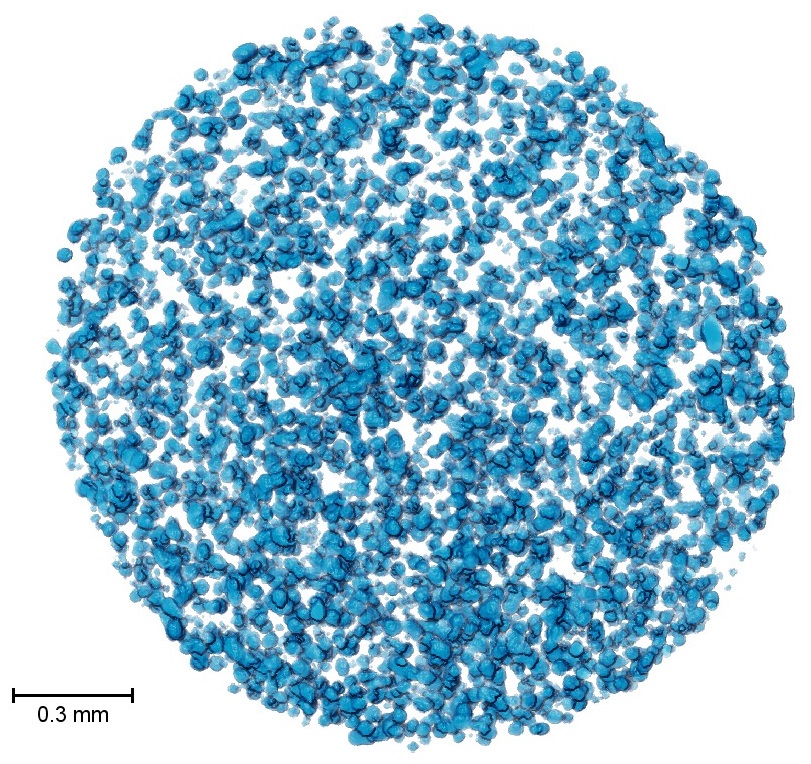}
	\caption{}
	\label{Particle2}
 \end{subfigure}
  \caption{Particle distribution examined using a CT scanner.}
  \label{AppendixB}
\end{figure}

Fig.~\ref{AppendixB}(a) displays a reconstructed 3D CT image with a vertical arrangement of the filament, where X-ray attenuation intensity is color-coded. Fig.~\ref{AppendixB}(b) presents a cross-section (\SI{0.5}{\mm} thickness) of the scan, without color coding. Both images demonstrate a homogeneous distribution of particles, without observable radial or axial gradients. While this examination covers only a short section of the filament, it is assumed that the properties remain consistent along its length.

\subsection{Rheometric data}
\label{Rheometric data}

\begin{figure}[H]
 \centering
 \begin{subfigure}[b]{.49\textwidth}
	\centering
		\includegraphics[width=\linewidth]{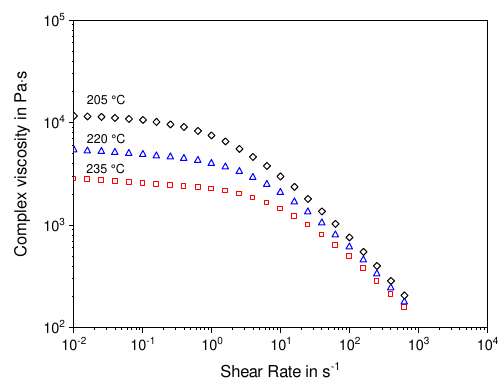}
	\caption{}
	\label{Wolfram}
 \end{subfigure}
 \begin{subfigure}[b]{.49\textwidth}
	\centering
		\includegraphics[width=\linewidth]{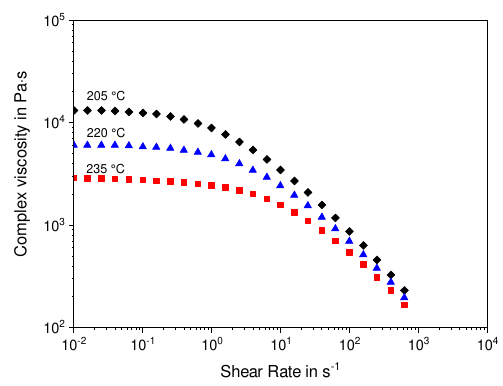}
	\caption{}
	\label{pure}
 \end{subfigure}
  \caption{Shear rate versus complex viscosity as a function of temperature for (a) tungsten powder filled polystyrene and (b) unfilled polystyrene.}
  \label{ShortLongTrans}
\end{figure}

To investigate the influence of metal powder addition and establish a basis for numerical simulations, dynamic viscosity tests were conducted on unfilled and tungsten powder-filled polystyrene. The results are depicted in Fig.~\ref{ShortLongTrans}.

\subsection{Flow field}
\label{Flow field}

\begin{figure}[h]
	\centering
		\includegraphics[scale=0.55]{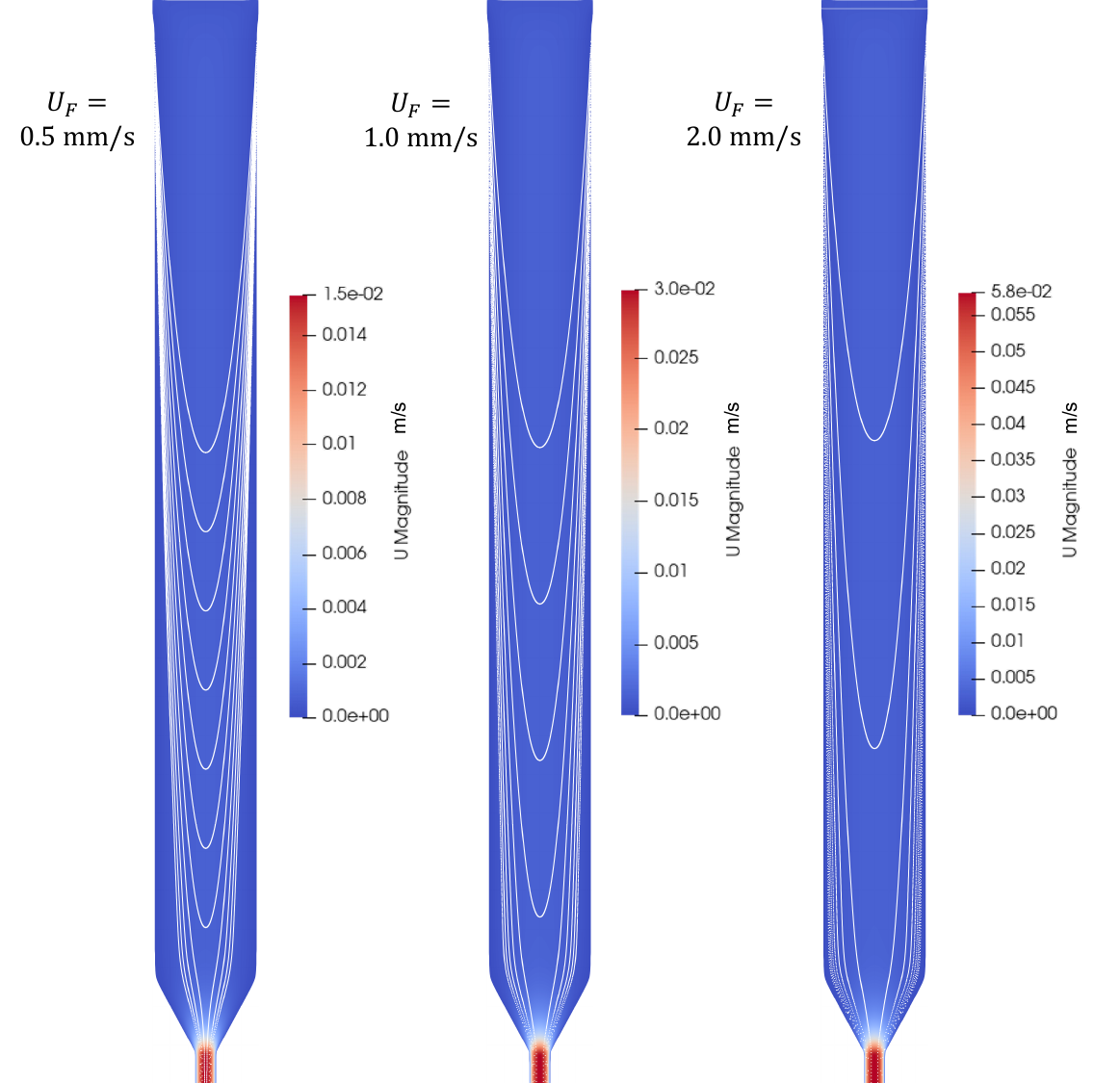}
	\caption{Numerically solved velocity field for three different filament speeds, along with calculated time lines for comparison with X-ray data.}\protect
	\label{Flow_field_figure}
\end{figure}

Fig.~\ref{Flow_field_figure} displays the numerically solved velocity field for three different filament velocities, along with computed time lines. These time lines, separated by an interval of 2.0 mm, were used for comparison with the X-ray data in Fig.~\ref{Fig11_projectional_radiagraphy}. By comparing the shape of these time lines across the various filament speeds, insights can be gained regarding the velocity distribution.

\clearpage
\printbibliography
\clearpage

\end{document}